# Introducing LETOR 4.0 Datasets

Tao Qin and Tie-Yan Liu

{taoqin,tyliu}@microsoft.com

Microsoft Research Asia

LETOR is a package of benchmark data sets for research on LEarning TO Rank, which contains standard features, relevance judgments, data partitioning, evaluation tools, and several baselines. Version 1.0 was released in April 2007. Version 2.0 was released in Dec. 2007. Version 3.0 was released in Dec. 2008. This version, 4.0, was released in July 2009. Very different from previous versions (V3.0 is an update based on V2.0 and V2.0 is an update based on V1.0), LETOR4.0 is a totally new release. It uses the Gov2 web page collection (~25M pages) and two query sets from Million Query track[1] of TREC 2007 and TREC 2008. We call the two query sets MQ2007 and MQ2008 for short. There are about 1700 queries in MQ2007 with labeled documents and about 800 queries in MQ2008 with labeled documents. If you have any questions or suggestions about the datasets, please kindly email us (letor@microsoft.com). Our goal is to make the dataset reliable and useful for the community.

## Datasets

LETOR4.0 contains 8 datasets for four ranking settings derived from the two query sets and the Gov2 web page collection. The 5-fold cross validation strategy is adopted and the 5-fold partitions are included in the package. In each fold, there are three subsets for learning: training set, validation set and testing set. The datasets can be downloaded at the LETOR 4.0 website[2].

| Setting | Datasets |
|---|---|
| Supervised ranking | MQ2007 |
|  | MQ2008 |
| Semi-supervised ranking | MQ2007-semi |
|  | MQ2008-semi |
| Rank aggregation | MQ2007-agg |

---

[1] http://ir.cis.udel.edu/million/index.html

[2] http://research.microsoft.com/en-us/um/beijing/projects/letor/letor4download.aspx

|  | MQ2008-agg |
|---|---|
| Listwise ranking | MQ2007-list |
|  | MQ2008-list |

**Descriptions**

- Supervised ranking  
  There are three versions for each dataset in this setting: NULL, MIN, QueryLevelNorm.
    - NULL version: Since some document may do not contain query terms, we use ``NULL'' to indicate language model features, for which would be minus infinity values. This version of the data cannot be directly be used for learning; the ``NULL'' should be processed first.
    - MIN version: Replace the ``NULL'' value in NULL version with the minimal vale of this feature under a same query. This data can be directly used for learning.
    - QueryLevelNorm version: Conduct query level normalization based on data in MIN version. This data can be directly used for learning. We further provide 5 fold partitions of this version for cross fold validation.

  Each row is a query-document pair. The first column is relevance label of this pair, the second column is query id, the following columns are features, and the end of the row is comment about the pair, including id of the document. The larger the relevance label, the more relevant the query-document pair. A query-document pair is represented by a 46-dimensional feature vector. Here are several example rows from MQ2007 dataset:  
  ==================================  
  2 qid:10032 1:0.056537 2:0.000000 3:0.666667 4:1.000000 5:0.067138 ... 45:0.000000 46:0.076923 #docid = GX029-35-5894638 inc = 0.0119881192468859 prob = 0.139842  
  0 qid:10032 1:0.279152 2:0.000000 3:0.000000 4:0.000000 5:0.279152 ... 45:0.250000 46:1.000000 #docid = GX030-77-6315042 inc = 1 prob = 0.341364  
  0 qid:10032 1:0.130742 2:0.000000 3:0.333333 4:0.000000 5:0.134276 ... 45:0.750000 46:1.000000 #docid = GX140-98-13566007 inc = 1 prob = 0.0701303  
  1 qid:10032 1:0.593640 2:1.000000 3:0.000000 4:0.000000 5:0.600707 ... 45:0.500000 46:0.000000 #docid = GX256-43-0740276 inc = 0.0136292023050293 prob = 0.400738  
  ==================================

- Semi-supervised ranking  
  The data format in this setting is the same as that in supervised ranking setting.

The only difference is that the datasets in this setting contains both judged and undged query-document pair (in training set but not in validation and testing set) while the datasets in supervised ranking contain only judged query-document pair. The relevance label ``-1'' indicates the query-document pair is not judged. An example is shown as follow.

```
==============================
-1 qid:18219 1:0.022594 2:0.000000 3:0.250000 4:0.166667 ... 45:0.004237 46:0.081600 #docid = GX004-66-12099765 inc = -1 prob = 0.223732
0 qid:18219 1:0.027615 2:0.500000 3:0.750000 4:0.333333 ... 45:0.010291 46:0.046400 #docid = GX004-93-7097963 inc = 0.0428115405134536 prob = 0.860366
-1 qid:18219 1:0.018410 2:0.000000 3:0.250000 4:0.166667 ... 45:0.003632 46:0.033600 #docid = GX005-04-11520874 inc = -1 prob = 0.0980801
==============================
```

- Rank aggregation

  In the setting, a query is associated with a set of input ranked lists. The task of rank aggregation is to output a better final ranked list by aggregating the multiple input lists. A row in the data indicate a query-document pair. Several rows are shown as below.

```
==============================
0 qid:10002 1:1 2:30 3:48 4:133 5:NULL ... 25:NULL #docid = GX008-86-4444840 inc = 1 prob = 0.086622
0 qid:10002 1:NULL 2:NULL 3:NULL 4:NULL 5:NULL ... 25:NULL #docid = GX037-06-11625428 inc = 0.0031586555555558 prob = 0.0897452
2 qid:10032 1:6 2:96 3:88 4:NULL 5:NULL ... 25:NULL #docid = GX029-35-5894638 inc = 0.0119881192468859 prob = 0.139842
==============================
```

  The first column is relevance label of this pair, the second column is query id, the following columns are ranks of the document in the input ranked lists, and the end of the row is comment about the pair, including id of the document.In the above example, 2:30 means that the ranks of the document is 30 in the second input list. Note that large ranks mean top positions in the input ranked list, and ``NULL'' means the document does not appear in a ranked list. The larger the relevance label, the more relevant the query-document pair. There are 21 input lists in MQ2007-agg dataset and 25 input lists in MQ2008-agg dataset.

- Listwise ranking

  The data format in the setting is very similar to that in supervised ranking. The difference is that the ground truth of this setting is a permutation for a query instead of multiple level relevance judgments. As shown in the following examples, the first column is the relevance degree of a document in ground truth permutation. Large value of the relevance degree means top position of the document in the permutation. The other columns are the same as that in the setting of supervised ranking.

```
==============================
```

```
1008 qid:10 1:0.004356 2:0.080000 3:0.036364 4:0.000000 ... 46:0.000000
#docid = GX057-59-4044939 inc = 1 prob = 0.698286
1007 qid:10 1:0.004901 2:0.000000 3:0.036364 4:0.333333 ... 46:0.000000
#docid = GX235-84-0891544 inc = 1 prob = 0.567746
1006 qid:10 1:0.019058 2:0.240000 3:0.072727 4:0.500000 ... 46:0.000000
#docid = GX016-48-5543459 inc = 1 prob = 0.775913
1005 qid:10 1:0.004901 2:0.160000 3:0.018182 4:0.666667 ... 46:0.000000
#docid = GX068-48-12934837 inc = 1 prob = 0.659932
==============================
```

## More low level information

In this section, we briefly introduce some raw data which can be used to extract new features. All the data described in this section can be downloaded at the LETOR 4.0 website.

- Meta data
  Meta data for all queries in the two query sets. The information can be used to reproduce some features like BM25 and LMIR, and can also be used to construct some new features.
    o Meta data for MQ2007 query set, ~ 60M
    o Meta data for MQ2008 query set, ~ 50
    o Collection info, ~1 k
- Link graph (~ 480M) of Gov2 collection
  Each line contains the inlinks of a web page. The first column is the MSRA doc id of the web page, the second column is the number of inlinks of this page, and the following columns list the MSRA doc ids of all the inlinks of this page. The mapping from MSRA doc id to TREC doc id can be found at the website.
- Sitemap (~ 65M) of Gov2 collection
  Each line is a web page. The first column is the MSRA doc id of the page, the second column is the depth of the url (number of slashes), the third column is the length of url (without ``http://"), the fourth column is the number of its child pages in the sitemap, the fifth column is the MSRA doc id of its parent page (-1 indicates no parent page).
- Similarity relation of Gov2 collection
  The data is organized by queries. The order of queries in the two files is the same as that in Large_null.txt in the MQ2007-semi dataset and MQ2008-semi dataset. The order of documents of a query in the two files is also the same as that in Large_null.txt in the MQ2007-semi dataset and MQ2008-semi dataset.
  Each row in the similarity files describes the similarity between a page and all the other pages under a same query. Note that i-th row in the similar files is exactly corresponding to the i-th row in Large_null.txt in MQ2007-semi dataset or MQ2008-semi dataset. Here is the an example line:
  ==========================

*qid:10002 qdid:1 406:0.785623 178:0.785519 481:0.784446 63:0.741556 882:0.512454 …*

============================

The first column shows the query id, and the second column shows the page index under the query. For example, for a query with 1000 web pages, the page index ranges from 1 to 1000. The following columns show the similarity between this page and the other pages. For example, 406:0.785623 indicates that the similarity between this page (with index 1 under the query) and the page (with index 406 under the query) is 0.785623. We sort the pages according to the descending order of similarity. The similarity between two pages is cosine similarity between the contents of the two pages.

### Feature list for supervised ranking

LETOR 4.0 supervised ranking datasets contain the following 46 features for learning to rank.

| Column in Output | Description |
| --- | --- |
| 1 | TF(Term frequency) of body |
| 2 | TF of anchor |
| 3 | TF of title |
| 4 | TF of URL |
| 5 | TF of whole document |
| 6 | IDF(Inverse document frequency) of body |
| 7 | IDF of anchor |
| 8 | IDF of title |
| 9 | IDF of URL |
| 10 | IDF of whole document |
| 11 | TF*IDF of body |
| 12 | TF*IDF of anchor |
| 13 | TF*IDF of title |
| 14 | TF*IDF of URL |
| 15 | TF*IDF of whole document |
| 16 | DL(Document length) of body |
| 17 | DL of anchor |
| 18 | DL of title |
| 19 | DL of URL |
| 20 | DL of whole document |
| 21 | BM25 of body |
| 22 | BM25 of anchor |
| 23 | BM25 of title |
| 24 | BM25 of URL |

| 25 | BM25 of whole document |
| --- | --- |
| 26 | LMIR.ABS of body |
| 27 | LMIR.ABS of anchor |
| 28 | LMIR.ABS of title |
| 29 | LMIR.ABS of URL |
| 30 | LMIR.ABS of whole document |
| 31 | LMIR.DIR of body |
| 32 | LMIR.DIR of anchor |
| 33 | LMIR.DIR of title |
| 34 | LMIR.DIR of URL |
| 35 | LMIR.DIR of whole document |
| 36 | LMIR.JM of body |
| 37 | LMIR.JM of anchor |
| 38 | LMIR.JM of title |
| 39 | LMIR.JM of URL |
| 40 | LMIR.JM of whole document |
| 41 | PageRank |
| 42 | Inlink number |
| 43 | Outlink number |
| 44 | Number of slash in URL |
| 45 | Length of URL |
| 46 | Number of child page |

## Acknowledgement

- The following people contributed to the construction of the LETOR4.0 dataset: Wenkui Ding, Jun Xu, Hang Li, Ben Carterette, Javed Aslam, James Allan, Stephen Robertson, Virgil Pavlu, and Emine Yilmaz .
- We would like to thank the following teams to kindly and generously share their runs submitted to TREC2007/2008: NEU team, U. Massachusetts team, I3S_Group_of_ICT team, ARSC team, IBM Haifa team, MPI-d5 team, Sabir.buckley team, HIT team, RMIT team, U. Amsterdam team, U. Melbourne team.